\documentclass[twocolumn,aps,prd,superscriptaddress,floatfix]{revtex4}
\usepackage{amssymb}
\usepackage{amsmath,bm}
\usepackage[normalem]{ulem}
\usepackage[dvips]{color}
\usepackage{booktabs}
\usepackage{threeparttable}
\usepackage{graphicx}
\usepackage{CJK}
\usepackage[colorlinks,
            linkcolor=blue,
            anchorcolor=blue,
            citecolor=blue]{hyperref}
\renewcommand{\sout}{\bgroup \color[rgb]{1,0,0}\ULdepth=-.5ex \ULset}

\begin{document}

\title{Equation of state of dense matter in the multimessenger era}

\author{Ying Zhou}
\affiliation{School of Physics and Astronomy and
Shanghai Key Laboratory for Particle Physics and Cosmology,
Shanghai Jiao Tong University, Shanghai 200240, China}
\author{Lie-Wen Chen\footnote{%
Corresponding author: lwchen$@$sjtu.edu.cn}}
\affiliation{School of Physics and Astronomy and
Shanghai Key Laboratory for Particle Physics and Cosmology,
Shanghai Jiao Tong University, Shanghai 200240, China}
\author{Zhen Zhang}
\affiliation{Sino-French Institute of Nuclear Engineering and Technology,
Sun Yat-sen University, Zhuhai 519082, China}

\date{\today}

\begin{abstract}
While the equation of state (EOS) of symmetric nuclear matter (SNM) at suprasaturation densities
has been relatively well constrained from heavy-ion collisions, the EOS of high-density neutron-rich
matter is still largely uncertain due to the poorly known
high-density behavior of the symmetry energy.
Using the constraints on the EOS of SNM at suprasaturation densities from
heavy-ion collisions together with the data of finite nuclei and the existence of $2M_\odot$ neutron
stars from electromagnetic (EM) observations, we show that the high-density symmetry energy cannot be
too soft, which leads to lower bounds on dimensionless tidal deformability of $\Lambda_{1.4} \ge 193$
and radius of $R_{1.4} \ge 11.1$ km for $1.4M_\odot$ neutron star.
Furthermore, we find that the recent constraint of $\Lambda_{1.4} \le 580$
from the gravitational wave signal GW170817 detected from the binary neutron
star merger by the LIGO and Virgo Collaborations rules out too stiff high-density
symmetry energy, leading to an upper limit of $R_{1.4} \le 13.3$ km.
All these terrestrial nuclear experiments and astrophysical observations based on strong, EM
and gravitational measurements together put stringent constraints on the high-density symmetry energy and
the EOS of SNM, pure neutron matter and neutron star matter.

\end{abstract}

\maketitle

\section{Introduction}
Dense matter with density comparable to nuclear saturation density $n_0$
($\sim 0.16$ nucleon/fm$^3 \approx 2.7 \times 10^{14}$ g/cm$^3$) can exist
in heavy atomic nuclei, in compact stars, or
be produced in heavy-ion collisions.
A basic model for understanding such dense matter is the nuclear matter -
an ideal static infinite uniform system composed of nucleons (neutrons and
protons) with only the strong interaction considered.
One fundamental issue in nuclear physics, particle physics and astrophysics
is to explore the equation of state (EOS) of nuclear
matter~\cite{Dan02,Lat04,Oer17}, conventionally defined as
energy (or pressure) vs density.
Because of the complicated nonperturbative feature of quantum chromodynamics
(QCD), it is still a big challenge to determine the nuclear matter EOS from
{\it ab initio} QCD calculations, especially at
suprasaturation densities~\cite{Bra14}.
Therefore, data from terrestrial experiments or
astrophysical observations are particularly important to constrain the nuclear
matter EOS.

Indeed, the EOS of symmetric nuclear matter (SNM) with equal fraction of neutrons
and protons has been relatively well constrained from around $n_0$ to about $5n_0$
by analyzing the data on giant monopole
resonance of heavy nuclei~\cite{You99,Gar18} as well as the kaon production~\cite{Aic85,Fuc06}
and collective flow~\cite{Dan02} in heavy-ion collisions.
On the other hand,
the EOS of dense neutron-rich matter, especially at suprasaturation densities,
remains largely uncertain due to the poorly known high-density behavior of
the isospin-dependent part of nuclear matter EOS,
characterized by the symmetry energy $E_{\rm sym}(n)$
(see, e.g., Ref.~\cite{LCK08}).

Nuclear data, including those from nuclear structure and heavy-ion collisions,
are usually difficult to constrain the high-density $E_{\rm sym}(n)$,
although its subsaturation density behavior has been relatively well determined.
For example, the nuclear mass can put stringent constraints on
$E_{\rm sym}(n)$ around $2/3n_0$ (the averaged density
of nuclei)~\cite{Zha13},
while the electric dipole polarizability is mainly sensitive to the
$E_{\rm sym}(n)$ around $1/3n_0$ since isovector giant dipole resonances are
essentially related to the neutrons and
protons in nuclear surface~\cite{Zha15}.
Heavy-ion collisions perhaps is the only way in terrestrial labs to produce high
density matter but the isospin asymmetry is usually small and thus
the current constraints on high-density $E_{\rm sym}(n)$ are
strongly model-dependent~\cite{Xiao09,Fen10,Rus11,Coz13,Rus16,Coz18}.

In nature,
neutron stars (NSs) provide an ideal site to explore dense matter.
The discovery of the currently heaviest neutron star
PSR J0348+0432~\cite{Ant13} with mass $2.01 \pm 0.04M_\odot$
actually rules out soft NS matter EOSs which are not stiff enough
against gravitational collapse.
The NS mass-radius ($M$-$R$) relation
has been shown to be sensitive to the high-density
$E_{\rm sym}(n)$~\cite{Lin92,Lat01,Oze10,Ste10} since the averaged density
of a NS is about $2.5n_0$ and the NS matter is dominated by
neutrons with a small fraction ($\sim 10\%$) of protons (and leptons to keep weak
equilibrium and charge neutrality).
Although the NS mass
can be determined precisely, the precise measurement of its radius remains a big
challenge~\cite{Oze16}.
A good probe of NS radius is the tidal deformability,
i.e., the ratio of the induced quadrupole moment of a neutron star
to the perturbing tidal field of its companion, and for a NS with mass $M$,
it can be expressed in dimensionless form as~\cite{Fla08,Hin08}
\begin{equation}
\Lambda_M=\frac{2}{3}k_2\bigg(\frac{c^2R}{GM}\bigg)^5,
\end{equation}
where $k_2$ is the tidal Love number and $R$ is the NS radius.
The inspiralling binary neutron star (BNS) merger,
one important source of gravitational wave (GW)
that can be detected by ground-based GW detectors,
provides a natural lab to extract information on $\Lambda_M$.
During the BNS inspiral stage before merger,
the tidal effects change the phase evolution of GW waveform
compared to that of a binary black hole (BBH) inspiral, and the difference
between BNS and BBH inspirals appears from the fifth post-Newtonian order
onwards with the leading order contribution proportional to
$\Lambda_M$~\cite{Fla08,Dam09,Gra18}.
The $\Lambda_M$ can thus be extracted from
the GW signal of BNS inspiral~\cite{Fla08,Hin10,Vin11,Dam12}.

On 17 August 2017, the first GW signal GW170817 of BNS merger
was observed and localized by the LIGO and Virgo
observatories~\cite{Abb17NSMerger},
and its electromagnetic (EM) radiation was also detected by many Collaborations
(see, e.g., Ref.~\cite{LAbb17EM}), inaugurating a new era of
multimessenger astronomy.
Using the GW170817 signal, a large number of
studies~\cite{Mar17,Bau17,Zho18,Fat18,De18,NBZ18,NBZ18b,Rad18,Cou18,Ann18,Mos18,Rez18,Rui18,Mal18}
have been performed to constrain the EOS of NS matter or the properties of NSs.
The original analysis of GW170817
suggests an upper limit of $\Lambda_{1.4} \le 800$~\cite{Abb17NSMerger},
and a more recent analysis~\cite{Abb18NSMerger} with some
plausible assumptions leads to a stronger constraint of
$\Lambda_{1.4} = 190^{+390}_{-120}$.

In this work, for the first time, by using the same model to
analyze simultaneously the data based on strong, EM
and gravitational measurements, i.e.,
the terrestrial data of finite nuclei and heavy-ion
collisions, the existence of $2M_{\odot}$
NS from EM observations and
the upper limit of $\Lambda_{1.4} \le 580$ from
GW170817, we put stringent constraints on
the high-density $E_{\rm sym}(n)$ and the EOS of SNM,
pure neutron matter (PNM) and NS matter.

\section{Methods}
The nuclear matter EOS, defined as the binding energy per nucleon,
can be expressed as the following parabolic approximation form
\begin{equation}\label{EOSANM}
E(n,\delta)=E_0(n)+E_{\mathrm{sym}}(n)\delta^2
+{\cal O}(\delta^4),
\end{equation}%
where $n=n_{\rm n}+n_{\rm p}$ is the nucleon number density
and $\delta=(n_{\rm n}-n_{\rm p})/n$ is the isospin asymmetry
with $n_{\rm p}$ and $n_{\rm n}$ denoting
the proton and neutron densities, respectively.
$E_0(n)=E(n,\delta=0)$ is the EOS of SNM,
and the symmetry energy is defined by
$E_{\mathrm{sym}}(n)=\left.\frac{1}{2!}\frac{\partial^{2}E(n,\delta)}{\partial\delta^{2}}\right|_{\delta=0}.$
At the saturation density $n_0$, the $E_0(n)$ can be expanded in
$\chi=(n-n_0)/3n_0$ as
$E_0(n)=E_0(n_0)+\frac{1}{2!}K_0\chi^2+\frac{1}{3!}J_0\chi^3+{\cal O}(\chi^4)$,
where $K_0$ is the incompressibility coefficient and $J_0$
is the skewness coefficient.
The $E_{\mathrm{sym}}(n)$ can be expanded at a reference density
$n_r$ in terms of the slope parameter $L(n_r)$ and the curvature
parameter $K_{\mathrm{sym}}(n_r)$ as
$E_{\mathrm{sym}}(n) = E_{\mathrm{sym}}(n_r) + L(n_r) \chi_r + \frac{1}{2!}K_{\mathrm{sym}}(n_r)\chi_r^2+\mathcal{O}(\chi_r^3)$,
with $\chi_r=(n-n_r)/(3n_r)$. Conventionally we have $L \equiv L(n_0)$ and
$K_{\mathrm{sym}} \equiv K_{\mathrm{sym}}(n_0)$.

In this work, we apply the extended Skyrme-Hartree-Fock
(eSHF) model~\cite{Cha09,Zha16} to three systems, i.e., nuclear matter, finite nuclei
and neutron stars.
As emphasized in Ref.~\cite{Zha16}, the
eSHF model includes additional momentum and density-dependent two-body forces
to effectively mimic the momentum dependence of the three-body force
and can very successfully describe
simultaneously the three systems which involve a wide
density region, and thus is especially suitable for our present
motivation. The extended Skyrme interaction is expressed
as~\cite{Cha09,Zha16}
\begin{eqnarray}\label{Eq:SHF}
v_{i,j}&=& t_0(1+x_0 P_\sigma)\delta(\bm{r})
    +\frac{1}{6}t_3(1+x_3 P_\sigma)
    n^\alpha(\bm{R}) \delta(\bm{r})\notag \\
&+& \frac{1}{2}t_1(1+x_1 P_\sigma)[K'^2\delta(\bm{r})
    +\delta(\bm{r})K^2]\notag\\
&+&t_2(1+x_2 P_\sigma)\bm{K}'\cdot
    \delta(\bm{r}){\bm{K}}\notag\\
&+&\frac{1}{2}t_4(1+x_4P_\sigma)
    [K'^2 \delta(\bm{r}) n(\bm{R})
    +n(\bm{R})\delta(\bm{r})  K^2] \notag\\
&+& t_5(1+x_5 P_\sigma)\bm{K}'\cdot
    n(\bm{R}) \delta(\bm{r}){\bm{K}} \notag \\
&+&iW_0(\bm{\sigma}_i+\bm{\sigma}_j)\cdot
    [\bm{K}'\times\delta(\bm{r}){\bm{K}}],
\end{eqnarray}
where the symbols have their conventional meaning~\cite{Cha09,Zha16}.
The interaction contains $14$ independent parameters,
i.e., the $13$ Skyrme parameters $\alpha$,
$t_0\sim t_5$, $x_0\sim x_5$,
and the spin-orbit coupling constant $W_0$.
Instead of directly using the $13$ Skyrme parameters,
we express them explicitly in terms of
the following $13$ macroscopic quantities (pseudo-parameters)~\cite{Zha16},
i.e., $n_0$, $E_{0}(n_0)$, $K_0$, $J_0$,
$E_{\rm sym}(n_r)$, $L(n_r)$, $K_{\rm sym}(n_r)$,
the isoscalar effective mass $m_{s,0}^{\ast}$,
the isovector effective mass $m_{v,0}^{\ast}$,
the gradient coefficient $G_S$,
the symmetry-gradient coefficient $G_V$,
the cross gradient coefficient $G_{SV}$,
and the Landau parameter $G_0'$ of SNM in the spin-isospin channel.
The higher-order parameters
$J_0$ and $K_{\rm sym}$
generally have small influence on the properties of finite nuclei
but are critical for the high-density neutron-rich matter EOS and NS properties.
In addition,
at the subsaturation density $n_c=0.11n_0/0.16$, the $E_{\rm sym}(n_c)$
has been precisely constrained to be $E_{\rm sym}(n_c) = 26.65\pm 0.2$ MeV~\cite{Zha13}
by analyzing the binding energy difference of heavy isotope pairs and
$L(n_c) = 47.3 \pm 7.8$ MeV~\cite{Zha14} is extracted from the electric dipole
polarizability of $^{208}$Pb.
Therefore, here we fix $J_0$ and $K_{\rm sym}$ at various values with
$E_{\rm sym}(n_c) = 26.65$ MeV and $L(n_c) = 47.3$ MeV,
and the other $10$ parameters are obtained by fitting
the data of finite nuclei
by minimizing the weighted sum of
the squared deviations between the theoretical predictions and the experimental data, i.e.,
$\chi^2(\bm{p})=\sum_{i=1}^{N}\left(\frac{{\cal O}_{i}^{\rm th}({\bm p})-{\cal O}_{i}^{\rm exp}}{\triangle {\cal O}_i}\right)^2,$
where the $\bm{p}=(p_1 ,...,p_z) $ define
the $z$ dimensional model space,
${\cal O}_i^{({\rm th})}$ and ${\cal O}_i^{({\rm exp})}$ are
the theoretical predictions and the corresponding experimental values of
observables, respectively, and $\Delta \mathcal{O}_i$ is the adopted error
used to balance the relative weights of the various
types of observables.
We note here that varying $L(n_c)$ within $L(n_c) = 47.3 \pm 7.8$ MeV mainly influences
the values of $G_S$, $G_V$ and $G_{SV}$ which are irrelevant to the nuclear matter EOS,
while the parameters $n_0$, $E_{0}(n_0)$, $K_0$ characterizing the nuclear matter EOS remain
almost unchanged.

In the fitting, we consider the following experimental data of
spherical even-even nuclei:
(i) the binding energies $E_{\rm{B}}$ of ${}^{16}$O, ${}^{40,48}$Ca,
${}^{56,68}$Ni, ${}^{88}$Sr, ${}^{90}$Zr, ${}^{100,116,132}$Sn,
${}^{144}$Sm and ${}^{208}$Pb \cite{Wan17};
(ii) the charge r.m.s. radii $r_{\rm{c}}$ of ${}^{16}$O, ${}^{40,48}$Ca,
${}^{56}$Ni, ${}^{88}$Sr, ${}^{90}$Zr, ${}^{116}$Sn, ${}^{144}$Sm and
${}^{208}$Pb \cite{Ang13,Fri95,Bla05};
(iii) the isoscalar giant monopole resonance energies $E_{\rm{GMR}}$
of ${}^{90}$Zr, ${}^{116}$Sn, ${}^{144}$Sm and ${}^{208}$Pb \cite{You99};
(iv) the spin-orbit energy level splittings $\epsilon_{\rm{ls}}^A$
for neutron $1p_{1/2}-1p_{3/2}$ and proton $1p_{1/2}-1p_{3/2}$ in ${}^{16}$O,
and the proton $2d_{3/2}-2d_{5/2}$, neutron $3p_{1/2}-3p_{3/2}$
and neutron $2f_{5/2}-2f_{7/2}$ in $^{208}$Pb \cite{Vau72}.
To balance the $\chi^2$ from each sort
of experimental data (see, e.g., Ref.~\cite{Zha16}), we assign the errors of $1.0$ MeV and $0.01$ fm to
the $E_{\rm{B}}$ and $r_{\rm{c}}$, respectively.
In particular, for the $E_{\rm{GMR}}$ we use the experimental error
multiplied by $3.5$ to also consider the impact of the experimental error,
while for the $\epsilon_{\rm{ls}}^A$ a $10\%$ relative error is employed.

For NSs, we consider here the conventional NS model
which includes only nucleons, electrons and possible muons ($npe\mu$),
and the NS is assumed to contain core, inner crust and outer
crust.
For the core, the EOS of $\beta$-stable and electrically neutral
$npe\mu$ matter is obtained from the eSHF model.
For the inner crust in the density region between
$n_{\rm out}$ and $n_{\rm t}$,
the EOS is constructed by interpolating with
$P=a+b{\cal E}^{4/3}$~\cite{Car03} where $P$ is pressure and $\cal E$ is energy density.
The density $n_{\rm{out}}$ separating the inner and the outer crusts
is taken to be $2.46 \times 10^{-4}\rm{fm}^{-3}$ while the core-crust transition
density $n_{\rm t}$ is evaluated self-consistently
by a dynamical approach~\cite{XuJ09a}.
For the outer crust, we use the well-known BPS EOS in the density
region of $6.93 \times 10^{-13} {\rm fm}^{-3}<n<n_{\rm out}$
and the FMT EOS for $4.73\times10^{-15}{\rm fm}^{-3}<n<6.93\times10^{-13}{\rm fm}^{-3}$~\cite{Bay71,Iid97}.
It should be noted that all the extended Skyrme interactions used in following
NS calculations satisfy the causality condition $dP/d{\cal E}<1$.

\begin{figure}[tbp]
\includegraphics[width=0.8\linewidth]{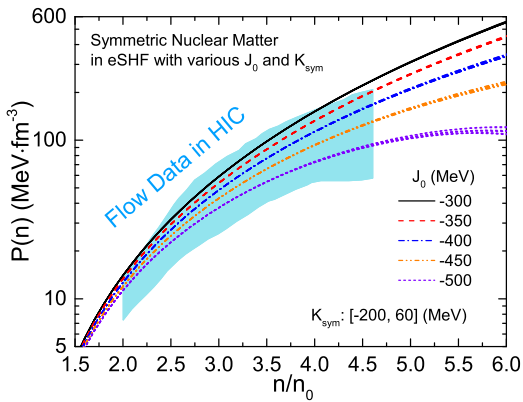}
\caption{(Color online) Pressure vs density for SNM
within eSHF in various extended Skyrme interactions with
$J_0$ and $K_{\rm sym}$ fixed at various values. The constraint from
collective flow data in heavy-ion collisions~\cite{Dan02} is
included for comparison.}
\label{fig1_Psnm-flow}
\end{figure}

\section{Results and Discussions}
Shown in Fig.~\ref{fig1_Psnm-flow} is pressure vs
density for SNM within eSHF in various extended Skyrme parameter sets
with $J_0$ fixed at ($-300$, $-350$, $-400$, $-450$, $-500$) MeV and
$K_{\rm sym}$ in the range of ($-200,60$) MeV. Also included in the
figure is the constraint from collective flow data in heavy-ion
collisions~\cite{Dan02}.
For all the parameter sets
with fixed $J_0$ and $K_{\rm sym}$, as expected, the total chi-square
$\chi_{\rm tot}^2$ falls in the range of $24.45<\chi_{\rm tot}^2<36.24$,
and the mean $\chi^2$ of each sort of experimental data
(i.e., $\chi^2_{E_{\rm B}}/12$, $\chi^2_{r_{\rm c}}/9$,
$\chi^2_{E_{\rm GMR}}/4$ and $\chi^2_{\epsilon_{\rm ls}^A}/5$)
is approximately equal to $1$.
The small variation of $\chi_{\rm tot}^2$ suggests that the higher-order
parameters $J_0$ and $K_{\rm sym}$ indeed have small influence
on the properties of finite nuclei.
In addition, the pressure of SNM exhibits negligible dependence on the
$K_{\rm sym}$, especially for $J_0 > -500$ MeV.
One sees that
the pressure of SNM becomes stiffer as the $J_0$ increases, and
$J_0 = -300$ MeV predicts a too stiff SNM EOS that violates
the flow data. A detailed study indicates an upper limit
at $J^{\rm up}_{0} = -342$ MeV.

\begin{figure}[tbp]
\includegraphics[width=0.9\linewidth]{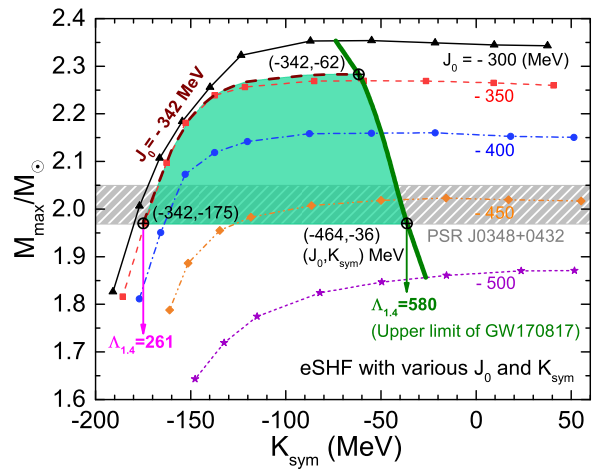}
\caption{(Color online) NS maximum mass $M_{\rm max}$ vs $K_{\rm sym}$
within eSHF in various extended Skyrme interactions with
$J_0$ and $K_{\rm sym}$ fixed at various values. See the text for details.}
\label{fig2_Mmax-J0Ksym}
\end{figure}

Using the same parameter sets as used in
Fig.~\ref{fig1_Psnm-flow}, we show
in Fig.~\ref{fig2_Mmax-J0Ksym} the NS maximum mass $M_{\rm max}$
vs $K_{\rm sym}$.
One sees the $M_{\rm max}$ increases sensitively
with increasing $J_0$ for a fixed $K_{\rm sym}$.
The $K_{\rm sym}$ has small influence on the $M_{\rm max}$ when
$K_{\rm sym}$ is greater than about $-100$ MeV,
but for a fixed $J_0$,
the $M_{\rm max}$ is drastically reduced with decreasing $K_{\rm sym}$ for
$K_{\rm sym} \lesssim -100$ MeV.
This behavior is due to the fact that a larger $K_{\rm sym}$ leads to
a stiffer high-density symmetry energy, and thus smaller
isospin asymmetry in high-density NS matter.
The smaller isospin asymmetry in turn suppresses
the sensitivity of $M_{\rm max}$ to $K_{\rm sym}$,
as the $M_{\rm max}$ is mainly determined by the EOS of high-density NS matter.
In particular, for sufficiently large $K_{\rm sym}$ (e.g., $\gtrsim -100$ MeV),
the NS matter can become almost isospin-symmetric at high densities,
and the NS maximum mass is therefore mainly sensitive to $J_0$
which dominates high-density behavior of symmetric nuclear matter.
Since the flow data in heavy-ion collisions require $J_0 \le -342$ MeV,
the parameter sets with $J_0 = -342$ MeV generally
predict largest values of $M_{\rm max}$, and the corresponding results
are also included Fig.~\ref{fig2_Mmax-J0Ksym}.
For $J_0 = -342$ MeV,
it is seen that when $K_{\rm sym}$ is smaller
than $-175$ MeV, the predicted $M_{\rm max}$ becomes violating
the mass lower limit (i.e., $1.97M_\odot$) of the heaviest NS
PSR J0348+0432~\cite{Ant13} observed so far (its mass $2.01 \pm 0.04M_\odot$
is shown as a shaded band in Fig.~\ref{fig2_Mmax-J0Ksym}), leading to a lower
limit at $K^{\rm low}_{\rm sym} = -175$ MeV.
For a fixed $J_0$, we find that the $\Lambda_{1.4}$
rapidly increases with increasing $K_{\rm sym}$.
For a fixed $K_{\rm sym}$, the $\Lambda_{1.4}$ also increases
with $J_0$ but much weaker than that
with $K_{\rm sym}$.
Our results indicate that the data of finite nuclei, the flow data
in heavy-ion collisions and the existence of $2M_\odot$ NS together
give the limit of $K_{\rm sym} \ge -175$ MeV, leading a
lower limit of $\Lambda^{\rm low}_{1.4} = 261$.

The above analyses mean the $E_{\rm sym}(n)$ cannot be too soft.
When the $K_{\rm sym}$ increases, the $E_{\rm sym}(n)$ becomes
stiffer and the $\Lambda_{1.4}$ increases accordingly.
The most recent
limit of $\Lambda_{1.4} \le 580$~\cite{Abb18NSMerger} thus can put an upper
limit for $K_{\rm sym}$ for each $J_0$ as indicated Fig.~\ref{fig2_Mmax-J0Ksym}.
The limit of $\Lambda_{1.4} \le 580$
together with the data of finite nuclei, the flow data in heavy-ion collisions
and the existence of $2M_\odot$ NS thus give an allowed region for the
higher-order parameters $J_0$ and $K_{\rm sym}$, as shown by green region in
Fig.~\ref{fig2_Mmax-J0Ksym}, which leads to
$-464~{\rm MeV} \le J_0 \le -342~{\rm MeV}$ and
$-175~{\rm MeV} \le K_{\rm sym} \le -36~{\rm MeV}$.
Moreover, the largest
NS mass is determined to be $2.28M_\odot$ at
$(J_0, K_{\rm sym}) = (-342, -62)$ MeV as indicated in Fig.~\ref{fig2_Mmax-J0Ksym}.

\begin{figure}[tbp]
\includegraphics[width=0.95\linewidth]{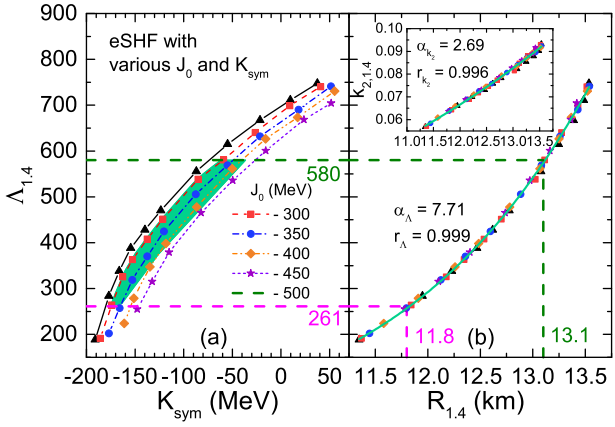}
\caption{(Color online) $\Lambda_{1.4}$ vs $K_{\rm sym}$ (a) and
$\Lambda_{1.4}$ vs $R_{1.4}$ (b) within eSHF using the same extended Skyrme
parameter sets as used in Fig.~\ref{fig2_Mmax-J0Ksym}.
The results for $k_{2,1.4}$ vs $R_{1.4}$ are included in the inset of
panel (b). See the text for details.}
\label{fig3_Lam14-Ksym-R14}
\end{figure}

Shown in Fig.~\ref{fig3_Lam14-Ksym-R14}(a) is
$\Lambda_{1.4}$ vs $K_{\rm sym}$ within eSHF using the same
parameter sets as used in Fig.~\ref{fig2_Mmax-J0Ksym}.
The corresponding
results for $\Lambda_{1.4}$ vs $R_{1.4}$ are shown in
Fig.~\ref{fig3_Lam14-Ksym-R14}(b) and
the inset in Fig.~\ref{fig3_Lam14-Ksym-R14}(b) displays  the results
for $k_{2,1.4}$ vs $R_{1.4}$.
The allowed region for $J_0$ and $K_{\rm sym}$ is also included in
Fig.~\ref{fig3_Lam14-Ksym-R14}(a).
As already mentioned, one indeed sees
that $\Lambda_{1.4}$ is sensitive to $K_{\rm sym}$ but is less affected by
the $J_0$.
From Fig.~\ref{fig3_Lam14-Ksym-R14}(b) and the inset,
one sees both $\Lambda_{1.4}$ and $k_{2,1.4}$ exhibit very strong correlation
with $R_{1.4}$, and they can be nicely fitted by the formulae
$\Lambda_{1.4}=a_\Lambda R_{1.4}^{\alpha_\Lambda}$ and
$k_{2,1.4}=a_{k_2} R_{1.4}^{\alpha_{k_2}}$, respectively, with
$a_\Lambda = (1.41\pm0.14)\times10^{-6}$, $\alpha_\Lambda = 7.71\pm0.04$,
$a_{k_2} = (8.25\pm0.58)\times10^{-5}$ and $\alpha_{k_2} = 2.69\pm0.03$.
The correlation coefficient is $r_\Lambda = 0.999$ for
$\Lambda_{1.4}=a_\Lambda R_{1.4}^{\alpha_\Lambda}$ and
$r_{k_2} = 0.996$ for $k_{2,1.4}=a_{k_2} R_{1.4}^{\alpha_{k_2}}$.
These relations together with $261 \le \Lambda_{1.4} \le 580$ lead to the
stringent constraints of
$R_{1.4}\in[11.8, 13.1]$ km and $k_{2,1.4}\in[0.064,0.085]$.

\begin{figure}[tbp]
\includegraphics[width=1.0\linewidth]{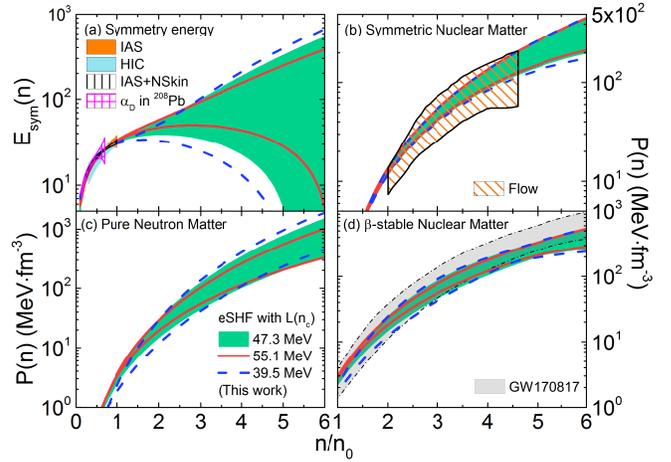}
\caption{(Color online) Density dependence of the symmetry energy (a) and
pressure for SNM (b), PNM (c) and neutron star matter (d).
See the text for details.}
\label{fig4_Esym-P}
\end{figure}

According to the allowed parameter space for
$J_0$ and $K_{\rm sym}$ as shown in Fig.~\ref{fig2_Mmax-J0Ksym},
we can determine the EOS of dense matter.
The obtained results for
$E_{\rm sym}(n)$ are shown in
Fig.~\ref{fig4_Esym-P}(a), and the pressure vs density for SNM, PNM
and NS matter is exhibited in Figs.~\ref{fig4_Esym-P}(b),
(c) and (d), respectively.
Also included in Fig.~\ref{fig4_Esym-P}(a) are
the constraints at subsaturation densities from
midperipheral heavy-ion collisions of Sn isotopes~\cite{Tsa09},
the isobaric analog states (IAS) and combining the neutron skin
data (IAS + NSkin)~\cite{Dan14}, and the electric dipole
polarizability ($\alpha_D$) in $^{208}$Pb \cite{Zha15}.
In addition, the constraints on pressure for SNM from flow data in heavy-ion
collisions~\cite{Dan02} and that for NS matter from GW170817~\cite{Abb18NSMerger}
are also included in Figs.~\ref{fig4_Esym-P}(b) and (d), respectively.
Furthermore, we include the corresponding results
with $L(n_c) = 55.1$ MeV and $39.5$ MeV to display the uncertainty
due to the $L(n_c)$ (i.e., $L(n_c)=47.3\pm7.8$ MeV~\cite{Zha14}).
One sees that our results are consistent with the existing constraints
but with much higher precision due to the simultaneous consideration
of the data of finite nuclei, the flow data in heavy-ion collisions, the
observed heaviest NS and the GW170817 signal.
We would like to point out that
the high density $E_{\rm sym}(n)$ still has large uncertainty and it could
be negative at high densities, which will cause isospin instability
and thus the presence of PNM core in NSs.

Furthermore, our results indicate that the $L(n_c) = 39.5$ MeV gives
smaller lower limits of $K^{\rm low}_{\rm sym}=-203$ MeV,
$\Lambda^{\rm low}_{1.4}=193$, $J^{\rm low}_{0}=-475$ MeV
and a heavier $M_{\rm max}=2.30M_\odot$,
while $L(n_c) = 55.1$ MeV gives lager lower limits of
$K^{\rm low}_{\rm sym}=-138$ MeV, $\Lambda^{\rm low}_{1.4}=380$,
$J^{\rm low}_{0}=-455$ MeV and a lighter $M_{\rm max}=2.26M_\odot$.
In addition,
the $E_{\rm sym}(2n_0)$ is found to be [46.9, 57.6] MeV,
[39.4, 54.5] MeV and [33.0, 51.3] MeV for $L(n_c) = 39.5$ MeV,
$47.3$ MeV and $51.1$ MeV, respectively.
Using $L(n_c)=47.3\pm7.8$ MeV,
therefore, we obtain
$J_0 \in [-464_{-11}^{+9}, -342]$ MeV,
$K_{\rm sym} \in [-175_{-28}^{+37}, -36\mp2]$ MeV,
$E_{\rm sym}(2n_0)\in[39.4_{+7.5}^{-6.4}, 54.5_{+3.1}^{-3.2}]$ MeV,
$\Lambda_{1.4} \in [261_{-68}^{+119}, 580]$,
$R_{1.4}\in[11.8_{-0.7}^{+0.8}, 13.1\pm0.2]$ km, and
$M_{\rm max} = 2.28\mp0.02~M_\odot$.

For the higher-order parameters $J_0$ and $K_{\rm sym}$,
the $J_0 \in[-464_{-11}^{+9},-342]$ MeV gives the strongest constraint
compared to the existing ones~\cite{Cai14}, and
the $K_{\rm sym} \in[-175_{-28}^{+37},-36\mp2]$ MeV is
also consistent with those extracted from the symmetry energy
systematics with some correlations~\cite{ChenLW11,ChenLW15,Mon17,Mal18}
or from heavy-ion collisions~\cite{Coz18}.
Moreover, the $E_{\rm sym}(2n_0)\in[39.4_{+7.5}^{-6.4}, 54.5_{+3.1}^{-3.2}]$ MeV
is in good agreement with those extracted from the symmetry energy
systematics~\cite{ChenLW15}, heavy-ion collisions~\cite{Rus16}
and the recent analyses on the NS observation and the
GW170817 signal~\cite{NBZ18,NBZ18b}.
As for the NS properties,
the $\Lambda^{\rm low}_{1.4} = 261_{-68}^{+119}$ agrees with the constraints
from analyzing the GW170817 signal~\cite{Ann18,Mos18} or its
EM signals~\cite{Rad18,Cou18},
the $R_{1.4}\in[11.8_{-0.7}^{+0.8}, 13.1\pm0.2]$ km agrees with those
from analyzing the GW170817~\cite{Ann18,Mos18,Mal18}, and the
$M_{\rm max} = 2.28\mp0.02~M_\odot$ is consistent with the results
from analyzing GW170817~\cite{Rez18,Rui18}.

Although the polytropic form of $P=a+b{\cal E}^{4/3}$ has been extensively
used to approximate the EOS of the NS inner crust~\cite{Lat01,Car03,XuJ09a,Zha16},
where the nuclear pasta may exist, the sensitivity of our results
to the choice of polytropic index needs to be studied.
As an example, based on the EOS with $J_0=-342$ MeV and $K_{\rm sym}=-175$ MeV,
which gives $M_{\rm max}= 1.97M_\odot$, we obtain
$\Lambda_{1.4}=261.76~(260.86, 260.50)$, $R_{1.4}=12.52~(11.82, 11.55)$ km
and $k_{2,1.4}=0.048~(0.064, 0.071)$
with the polytropic index $1~(4/3, 5/3)$.
These results indicate that the increase of the polytropic index leads to
considerable increase of the $k_{2,1.4}$, small reduction ($<10\%$) of $R_{1.4}$,
and negligible decrease of $\Lambda_{1.4}$, suggesting the choice of the polytropic
index has negligible effects on $\Lambda_{1.4}$.

\section{Summary}
Using the eSHF model to simultaneously analyze the data from
terrestrial nuclear experiments and astrophysical observations
based on strong, EM and gravitational measurements,
we have put stringent constraints on the high-density $E_{\rm sym}(n)$
and the pressure of SNM, PNM and NS matter.
We have found that the nuclear data and the existence of
$2M_{\odot}$ NS rule out too soft high-density $E_{\rm sym}(n)$,
leading to lower limits of $\Lambda_{1.4} \ge 193$ and
$R_{1.4} \ge 11.1$ km.
Further combining the upper limit of
$\Lambda_{1.4} \le 580$ from GW170817 excludes too stiff
high-density $E_{\rm sym}(n)$, leading to an upper limit
of $R_{1.4} \le 13.3$ km.
Using $L(n_c)=47.3\pm7.8$ MeV, we have obtained
$J_0 \in[-464_{-11}^{+9},-342]$ MeV,
$K_{\rm sym} \in[-175_{-28}^{+37},-36\mp2]$ MeV,
$E_{\rm sym}(2n_0)\in[39.4_{+7.5}^{-6.4}, 54.5_{+3.1}^{-3.2}]$ MeV,
and $M_{\rm max} = 2.28\mp 0.02~M_\odot$.
In future,
more precise limit on $L(n_c)$, possible discovery of heavier NS and
tighter bound on $\Lambda_{1.4}$ from BNS merger would put stronger
constraints on the high-density $E_{\rm sym}(n)$ and
thus the EOS of dense neutron-rich matter.

\section*{Acknowledgments}
The authors thank Tanja Hinderer, Ang Li, Bao-An Li and Jorge Piekarewicz
for helpful discussions.
This work was supported in part by the National Natural Science
Foundation of China under Grant No. 11625521, the Major State Basic Research
Development Program (973 Program) in China under Contract No.
2015CB856904, the Program for Professor of Special Appointment (Eastern
Scholar) at Shanghai Institutions of Higher Learning, Key Laboratory
for Particle Physics, Astrophysics and Cosmology, Ministry of
Education, China, and the Science and Technology Commission of
Shanghai Municipality (11DZ2260700).



\nocite{*}

\begin{thebibliography}{99}
\bibitem{Dan02} P. Danielewicz, R. Lacey, and W.G. Lynch,
        {Science \textbf{298}, 1592 (2002).}

\bibitem{Lat04} J.M. Lattimer and M. Prakash,
        {Science \textbf{304}, 536 (2004)}.

\bibitem{Oer17} M. Oertel, M. Hempel, T. Klahn, and S. Typel,
        {Rev. Mod. Phys. \textbf{89}, 015007 (2017)}.

\bibitem{Bra14} N. Brambilla {\it et al.},
        {Eur. Phys. J. C \textbf{74}, 2981 (2014).}

\bibitem{You99} D.H. Youngblood, H.L. Clark, and Y.-W. Lui,
        {Phys. Rev. Lett. \textbf{82}, 691 (1999).}

\bibitem{Gar18} U. Garg and G. Colo,
        {Prog. Phys. Nucl. Phys. \textbf{101}, 55 (2018).}

\bibitem{Aic85} J. Aichelin and C.M. Ko,
        {Phys. Rev. Lett. \textbf{55}, 2661 (1985).}

\bibitem{Fuc06} C. Fuchs,
        {Prog. Part. Nucl. Phys. \textbf{56}, 1 (2006).}

\bibitem{LCK08} B.A. Li, L.W. Chen, and C.M. Ko,
        {Phys. Rep. \textbf{464}, 113 (2008).}

\bibitem{Zha13} Z. Zhang and L.W. Chen,
        {Phys. Lett. B \textbf{726}, 234 (2013).}

\bibitem{Zha15} Z. Zhang and L.W. Chen,
        {Phys. Rev. C \textbf{92}, 031301(R) (2015).}

\bibitem{Xiao09} Z. Xiao, B.A. Li, L.W. Chen, G.C. Yong, and M. Zhang,
        {Phys. Rev. Lett. \textbf{102}, 062502 (2009).}

\bibitem{Fen10} Z.Q. Feng and G.M. Jin,
        {Phys. Lett. B \textbf{683}, 140 (2010).}

\bibitem{Rus11} P. Russotto {\it et al.},
        {Phys. Lett. B \textbf{697}, 471 (2011).}

\bibitem{Coz13} M.D. Cozma, Y. Leifels, W. Trautmann, Q. Li, and P. Russotto,
        {Phys. Rev. C \textbf{88}, 044912 (2013).}

\bibitem{Rus16} P. Russotto {\it et al.},
        {Phys. Rev. C \textbf{94}, 034608 (2016).}

\bibitem{Coz18} M.D. Cozma,
        {Eur. Phys. J. A \textbf{54}, 40 (2018).}

\bibitem{Ant13} J. Antoniadis {\it et al.},
        {Science \textbf{340}, 1233232 (2013).}

\bibitem{Lin92} L. Lindblom,
        {Astrophys. J. \textbf{398}, 569 (1992).}

\bibitem{Lat01} J.M. Lattimer and M. Parakash,
        {Astrophys. J. \textbf{550}, 426 (2001).}

\bibitem{Oze10} F. Ozel, G. Baym, and T. Guver,
        {Phys. Rev. D \textbf{82}, 101301(R) (2010).}

\bibitem{Ste10} A.W. Steiner, J.M. Lattimer, and E.F. Brown,
        {Astrophys. J. \textbf{722}, 33 (2010).}

\bibitem{Oze16} F. Ozel and P. Freire,
        {Annu. Rev. Astron. Astrophys. \textbf{54}, 401 (2016).}

\bibitem{Hin08} T. Hinderer,
        {Astrophys. J. \textbf{677}, 1216 (2008).}

\bibitem{Fla08} \'E.\'E. Flanagan and T. Hinderer,
        {Phys. Rev. D \textbf{77}, 021502(R) (2008).}

\bibitem{Dam09} T. Damour and A. Nagar,
        {Phys. Rev. D \textbf{80}, 084035 (2009).}

\bibitem{Gra18} S. E. Gralla,
        {Class. Quantum Grav. \textbf{35}, 085002 (2018).}

\bibitem{Hin10} T. Hinderer, B.D. Lackey, R.N. Lang, and J.S. Read,
        {Phys. Rev. C \textbf{81}, 123016 (2010).}

\bibitem{Vin11} J. Vines, \'E.\'E. Flanagan, and T. Hinderer,
        {Phys. Rev. D \textbf{83}, 084051 (2011).}

\bibitem{Dam12} T. Damour, A. Nagar, and L. Villain,
        {Phys. Rev. D \textbf{85}, 123007 (2012).}

\bibitem{Abb17NSMerger} B.P. Abbott \textit{et al.},
        {Phys. Rev. Lett. \textbf{119}, 161101 (2017).}

\bibitem{LAbb17EM} B.P. Abbott \textit{et al.},
        {Astrophys. J. Lett. \textbf{848}, L12 (2017).}

\bibitem{Mar17} B. Margalit and B.D. Metzger,
        {Astrophys. J. Lett. \textbf{850}, L19 (2017).}

\bibitem{Bau17} A. Bauswein, O. Just, H.-T. Janka, and N. Stergioulas,
        {Astrophys. J. Lett. \textbf{850}, L34 (2017).}

\bibitem{Zho18} E.-P. Zhou, X. Zhou, and A. Li,
        {Phys. Rev. D \textbf{97}, 083015 (2018).}

\bibitem{Fat18} F.J. Fattoyev, J. Piekarewicz, and C.J. Horowitz,
        {Phys. Rev. Lett. \textbf{120}, 172702 (2018).}

\bibitem{De18} S. De, D. Finstad, J.M. Lattimer, D.A. Brown, E. Berger, and C.M. Biwer,
        {Phys. Rev. Lett. \textbf{121}, 091102 (2018).}

\bibitem{NBZ18} N.B. Zhang, B.A. Li, and J. Xu,
        {Astrophys. J. \textbf{859}, 90 (2018).}

\bibitem{NBZ18b} N.B. Zhang and B.A. Li,
        {arXiv:1807.07698.}

\bibitem{Ann18} E. Annala, T. Gorda, A. Kurkela, and A. Vuorinen,
        {Phys. Rev. Lett. \textbf{120}, 172703 (2018).}

\bibitem{Mos18} E.R. Most, L.R. Weih, L. Rezzolla, and J. Schaffner-Bielich,
        {Phys. Rev. Lett. \textbf{120}, 261103 (2018).}

\bibitem{Rad18} D. Radice, A. Perego, F. Zappa, and S. Bernuzzi,
        {Astrophys. J. Lett. \textbf{852}, L29 (2018).}

\bibitem{Cou18} M. W. Coughlin {\it et al.},
        {arXiv:1805.09371.}

\bibitem{Rez18} L. Rezzolla, E.R. Most, and L.R. Weih,
        {Astrophys. J. Lett. \textbf{852}, L25 (2018).}

\bibitem{Rui18} M. Ruiz, S.L. Shapiro, and A. Tsokaros,
        {Phys. Rev. D \textbf{97}, 021501(R) (2018).}

\bibitem{Mal18} T. Malik {\it et al.},
        {Phys. Rev. C \textbf{98}, 035804 (2018).}

\bibitem{Abb18NSMerger} B.P. Abbott \textit{et al.},
        {Phys. Rev. Lett. \textbf{121}, 161101 (2018).}

\bibitem{Cha09} N. Chamel, S. Goriely, and J.M. Pearson,
        {Phys. Rev. C \textbf{80}, 065804 (2009).}

\bibitem{Zha16} Z. Zhang and L.W. Chen,
        {Phys. Rev. C \textbf{94}, 064326 (2016).}

\bibitem{Zha14} Z. Zhang and L.W. Chen,
        {Phys. Rev. C \textbf{90}, 064317 (2014).}

\bibitem{Wan17} M. Wang, G. Audi, F.G. Kondev, W.J. Huang, S. Naimi, and X. Xu
        {Chin. Phys. C \textbf{341}, 030003 (2017).}

\bibitem{Ang13} I. Angeli and K.P. Marinova,
        {At. Data Nucl. Data Tables \textbf{99}, 69 (2013).}

\bibitem{Fri95} G. Fricke, C. Bernhardt, K. Heilig, L.A. Schaller, L. Schellenberg, E.B. Shera, and C.W. Dejager,
         {At. Data Nucl. Data Tables \textbf{60}, 177 (1995)}.

\bibitem{Bla05} F. Le Blanc \textit{et al}.,
        {Phys. Rev. C \textbf{72}, 034305 (2005).}

\bibitem{Vau72} D. Vautherin, and D.M. Brink,
        {Phys. Rev. C \textbf{5}, 626 (1972).}

\bibitem{Car03} J. Carrier, C.J. Horowitz, and J. Piekarewicz,
        {Astrophys. J. \textbf{593}, 463 (2003).}

\bibitem{XuJ09a} J. Xu, L.W. Chen, B.A. Li, and H.R. Ma,
        {Astrophys. J. \textbf{697}, 1549 (2009).}

\bibitem{Bay71} G. Baym, C. Pethick, and P. Sutherland,
        {Astrophys. J. \textbf{170}, 299 (1971).}

\bibitem{Iid97} K. Iida and K. Sato,
        {Astrophys. J. \textbf{477}, 294 (1997).}

\bibitem{Tsa09} M. B. Tsang \textit{et al.},
        {Phys. Rev. Lett. \textbf{102}, 122701 (2009).}

\bibitem{Dan14} P. Danielewicz and J. Lee,
         {Nucl. Phys. A 922, 1 (2014).}

\bibitem{Cai14} B.J. Cai and L.W. Chen,
        {Nucl. Sci. Tech. \textbf{28}, 185 (2017).}

\bibitem{ChenLW11} L.W. Chen,
        {Sci. China Phys. Mech. Astron. \textbf{54}, suppl.1, s124 (2011).}

\bibitem{ChenLW15} L.W. Chen,
        {EPJ Web of Conferences \textbf{88}, 00017 (2015).}

\bibitem{Mon17} C. Mondal {\it et al.},
        {Phys. Rev. C \textbf{96}, 021302(R) (2017).}

\end{thebibliography}

\end{document}